\documentclass[]{spie}  

\usepackage{amsmath,amsfonts,amssymb}
\usepackage{graphicx}
\usepackage[colorlinks=true, allcolors=blue]{hyperref}

\title{3U Transat: a cubesat constellation to boost the multi-messenger astronomy }{}

\author[a]{Jean-Yves H{\'e}loret}
\author[a]{Olivier Godet}
\author[a]{Laurent Bouchet}
\author[a]{Jean-Luc Atteia}
\author[a]{Guillaume Orttner}
\affil[a]{IRAP, Universit{\'e} de Toulouse, CNRS, CNES, 9 Avenue du Colonel Roche, 31028 Toulouse, France}

\begin{document} 
\maketitle

\begin{abstract}
Thanks to the advent of sensitive gravitational wave (GW) and neutrino  detectors, multi-messenger (MM) astronomy will deeply transform our understanding of the Universe contents and evolution over cosmological times. To fully exploit the forthcoming GW and neutrino discoveries, it is crucial to detect as many electromagnetic (EM) counterparts as possible, but up to now, only one event has been detected by both GW detectors (Ligo/Virgo) and electromagnetic detectors (Fermi/GBM (Gamma ray Burst Monitor) and Integral), the  short gamma-ray burst GRB 170817A/GW 170817 associated with the merger of a binary neutron star.
 
To help improving the rate of joint MM events, it is crucial for the EM detectors in particular at high-energy in space to observe all the sky with a decent sensitivity. To do so, we propose the development of 3U Transat (TRANsient sky SATellites) project. 3U Transat is a constellation of nano-satellites offering a full sky coverage with a limited investment. The goal of this article is to present the 3U Transat project and its main scientific drivers as well as its current status. We will also describe our dynamic simulator used to optimise the scientific performances of the constellation. We will show highlights of the expected performances in term of detection and localisation capabilities as a function of the number of satellites in the constellation.
  
\end{abstract}


\section{INTRODUCTION}
\label{sec:intro}  

\begin{figure}[ht]
\centering
\includegraphics[scale = 0.5]{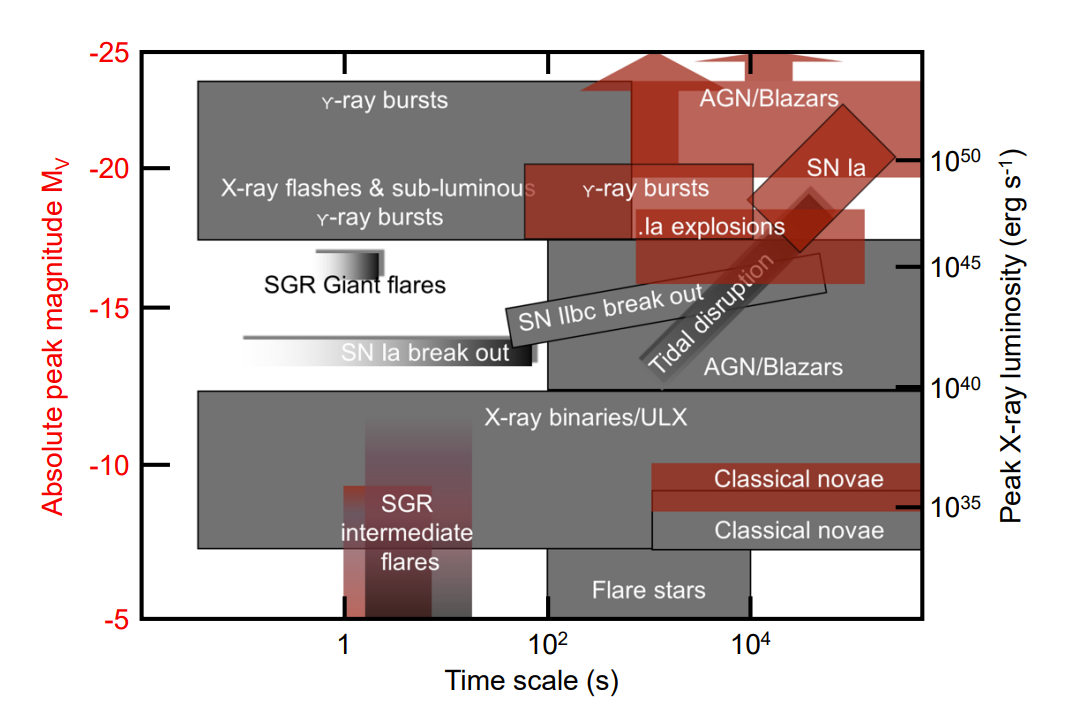}
\caption{: Observed transient phenomena 
				Credit: J. Wei et al., 2016 \cite{ref9}
}\label{Figure 1}
\end{figure}

Time-domain astronomy is pairing with multi-messenger (MM) astronomy. It is a forefront and rapidly rising branch of modern astronomy that will deeply transform our understanding of the evolution of the Universe and its contents in forthcoming years with the advent of more and more sensitive facilities (e.g. Vera Rubin Observatory \cite{ref1}, Cherenkov Telescope Array \cite{ref2}, Square Kilometre Array \cite{ref3} on the electromagnetic spectrum, Einstein telescope \cite{ref4}/Cosmic Explorer\cite{ref5} /Laser Interofometer Space Antenna \cite{ref6} for gravitational waves (GW), and IceCube \cite{ref7}/KM3Net \cite {ref8} for neutrinos).  Figure \ref{Figure 1} displays some of the most luminous compact object-driven transient phenomena known to release a huge amount of energy into their environment either as  radiation, kinetic energy, gravitational waves and/or particles.  In addition of giving us clues on how matter behaves under extreme conditions, the cosmological demography of various compact object populations and the compact object role in shaping their surrounding, these transients could also be used as cosmological tracers to study the early Universe and to measure cosmological parameters.

 Among these transient phenomena, Gamma-Ray Bursts (GRB) are the brightest ones \cite{ref10}, with an isotropic bolometric energy from a few $10^{46}$ ergs up to $10^{55}$ ergs, while being observable over cosmological distances up to redshift 9.3 \cite{ref11}. GRBs are characterised by a prompt emission in X-/$\gamma$-rays lasting from a few ms to a few hundreds of seconds followed by a long-lived multi-wavelength emission\cite{ref12}. GRBs signal the violent birth of a black-hole (BH) or a highly magnetised neutron star (NS) and the launch of powerful ultra-relativistic jets more or less oriented towards the Earth. GRBs are usually classified into two categories: short GRBs with a prompt emission lasting less than 2 seconds and long GRBs with a prompt emission lasting more than 2 seconds. Short GRBs are in most cases due to the merger of NS or NS-BH binary, while long GRBs are in most cases due to the gravitational core-collapse of some metal-poor and rapidly rotating massive stars \cite{ref12}. However, it has been shown that this classification presents some exceptions \cite{ref13}.

To fully exploit the wealth of forthcoming discoveries from MM facilities, it is crucial to detect as many electromagnetic (EM) counterparts of MM events as possible \cite{ref13a}. However despite thorough EM searches following GW and neutrino events, only one such event has been detected so far: the GW event GW\,170817 detected by Ligo and signalling the merger of a binary NS, followed a few seconds later by a short, underluminous and off-axis Gamma-Ray Burst GRB\,170817A and a few hours later by a r-process nucleosynthesis-driven kilonova detected in optical \cite{ref14}.

This lack of EM counterparts is partly because while GW and neutrino detectors observe all the sky when operating, current EM instrumentation, in particular at high-energy (HE), only cover small fractions of the sky at a given time; it is also due to the rather poor localising capabilities of current GW and neutrino facilities (typically 100 to 1000 $deg^{2}$); the third reason is that the relativistic jets are not necessarily oriented towards the Earth. When it is the case, the prompt emission is underluminous and more difficult to detect at large distances ($> 100 Mpc$) with current HE facilities. 
However, as shown by the study of GRBs, detecting prompt HE emission is often the key to detect counterparts at other wavelengths and to assess the true energetic of transient events. Also note that the follow-up ground segment instrumentation in NIR/optical has tried to adapt through the development of world networks of robotic wide-field telescopes \cite{ref15} allowing prompt tilling of large portions of the sky through sophisticated and coordinated observing strategies.  

In order to bridge this performance discrepancy, we need to design adapted HE instrumentation. How this will be done in term of instrumentation and observing strategy in forthcoming years is still unclear. This may rely on the development of a large mission with HE prompt and multi-wavelength follow-up observing capabilities like the Neil Gehrels \textit{Swift} observatory \cite{ref16}, SVOM (Space Variable Object Monitor) \cite{ref17} or Theseus (Transient High Energy Sky and Early Universe Surveyor) \cite{ref18} and/or constellations of micro/nano-satellites (e.g. Hermes \cite{ref19}, Camelot(Cubesat Applied for MEasuring and LOcalising Transient) \cite{ref20}, CATCH (Chasing All Transient Constellation Hunter) space missions \cite{ref21}). In any case, this new HE instrumentation shall have all-sky monitoring capabilities and be sufficiently sensitive to detect underluminous, nearby ($< 100$ Mpc) and rare events like GRB\,170817A.

Several researchers at IRAP (Institut de Recherche en Astrophysique et Plan\'etologie) proposed to the French Space Agency (CNES-Centre National d'Etudes Spatiales) to develop a Low Earth Orbit (LEO) constellation of cubesats in a 3U (3 litres) format called 3U Transat to survey the transient sky in synergy with MM facilities \cite{ref22}. It is a two step project: the first one being to develop a 3 yr demonstrator made of three identical cubesats to work in synergy with ground GW detectors during the O5 run (currently starting mid-2027) as well as with the SVOM mission. The demonstrator is designed to prove in-orbit the concept feasibility, to prepare the technical choices and the collaborations for the operational phase. The second step will deal with the cubesat building and deployment strategy of the full constellation at later times. 
 
We have conducted a phase 0 for the project from March 2020 to September 2022 with the help of CNES and U-Space (a company designing and building nano-satellites in Toulouse, France). This successful phase 0 has demonstrated the viability of the 3U Transat concept. Presently, we develop a full prototype of the science payload to increase its TRL (Technical Readiness Level) up to 5 minimum. Once completed in 2025, the performances of this prototype will be fully investigated using our lab experimental facility. 

This article is organised as follows: in Section \ref{sec2}, we present the 3U Transat project, in particular its main science drivers as well as the description of the science payload. In Section \ref{sec3} we present the end-to-end simulator of 3U Transat used to evaluate the performances of the constellation. In Section \ref{sec4}, we present some performance highlights of the constellation as a function of the number of cubesats considered. Section \ref{sec5} is devoted to the discussion and concluding remarks.

\section{3U Transat}\label{sec2}
\subsection{Main science drivers}\label{sec2.1}

The main objective of the 3U Transat project is to provide an instrumentation offering more than 80\% sky coverage and able to detect underluminous GRBs like GRB\,170817A (and potentially other transient phenomena) in hard X-rays up to more or less 60 Mpc. 

 The second objective is to allow alert broadcasting to the follow-up and MM instruments in less than 4-5 hours in total, i.e. from the detection of a GRB signal on board up to the issue of the alert. This delay is expected to be sufficient to allow the follow-up instruments to detect and observe the afterglow and/or the kilonova in case of the detection of off-axis bursts. 

The third objective is that 3U Transat is able to localise GRBs with a median error box (at 90\% confidence level) of less than 60 $deg^{2}$ and, for bright GRBs, with a median 90\% c.l. error box of less than 10 $deg^{2}$.

The fourth objective is to participate to the world effort to detect HE counterparts of MM events by finding means to combine data from different instruments similar to 3U Transat to improve both detection and localisation capabilities with the ultimate goal to develop a collaborative data analysis ground segment. We anticipate to make use of SVOM data as a first benchmark to do so. 
 
Thanks to this project, we also plan to investigate the spatialization of silicon photo-multipliers (SiPM) coupled with inorganic scintillators for HE astrophysical applications (see Section \ref{descrip}).

The lifetime of the demonstrator is requested to be of 3 years, while for the whole constellation we request the lifetime to be longer than 4 years. During these periods, 3U Transat is requested to detect and localise N short GRBs per year: N $> 2-3$ for the demonstration phase (3 satellites); N $> 20$ for the whole constellation.

\subsection{Overall description} \label{descrip}

The 3U Transat project consists of a constellation of identical 3U nano-satellites in LEO (Low Earth Orbit) and of a Ground Segment, itself composed of an operational part and of a scientific part. The total number of satellites is not fixed yet and will be derived from the performance simulation allowing to reach our scientific drivers.\
 
\begin{figure}[ht]
\centering
\includegraphics[scale = 0.6]{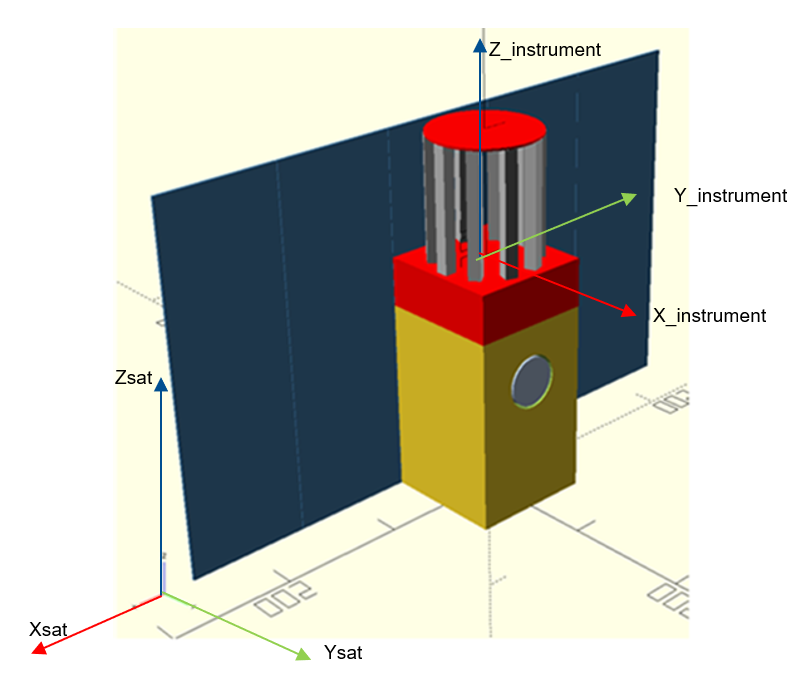}
\caption{: Satellite axis. 
				Credit: CNES
}\label{Figure 2}
\end{figure}

The Phase 0 studies showed that adopting a geocentric pointing attitude with yaw-steering orientation, i.e. Z axis pointing at the zenith, X axis toward the radial direction and Y axis toward the Sun as much as possible (see Figure \ref{Figure 2} for the definition of satellite and payload axis) is a convenient way to have varied pointing directions as well as ensuring a good illumination of the solar panels along the orbit.\ 

The constellation will be deployed in LEO with an altitude between 500 and 600 km. For the demonstration phase, the 3 satellites will be likely deployed on the same Sun Synchronous Orbit (SSO) because their launch is anticipated to be in a piggy bag of a scientific or commercial mission. Note that SSO orbits are not optimal since the cubesats will pass regularly in the polar cusps as well as in the South Atlantic Anomaly (SAA) reducing the observing duty cycle of each cubesat as well as inducing an ageing of the SiPM and electronics that may degrade the science performances of the payload during the mission. 

\begin{figure}[ht]
\centering
\includegraphics[scale = 0.4]{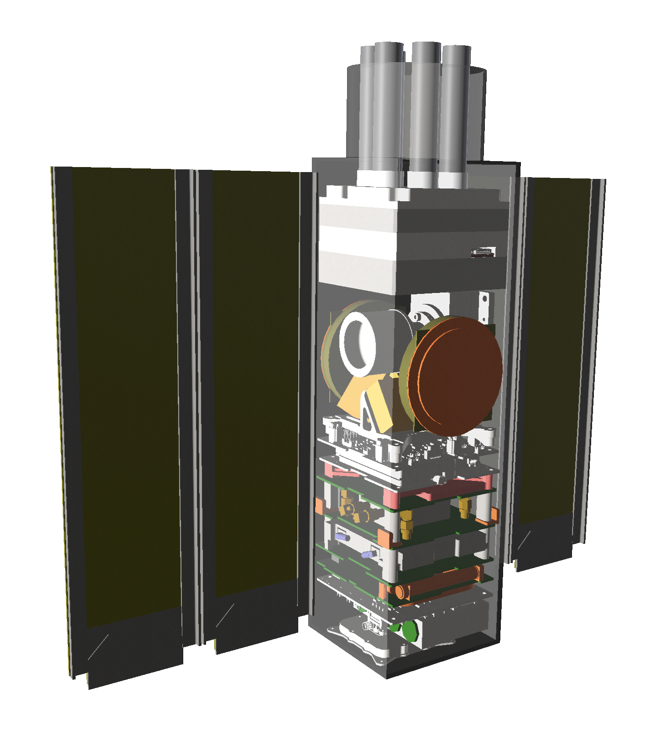}
\includegraphics[scale = 0.4]{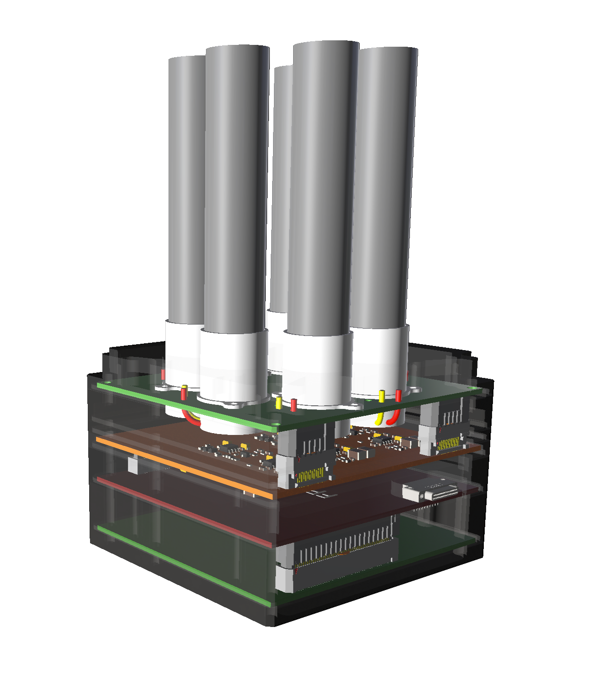}
\caption{: Overview of a 3U-Transat satellite (left) and of the payload (right)
}\label{Figure 3}
\end{figure}

Each satellite is designed to fit with the 3U format (1U is equivalent to 1 $dm^{3} $). The science payload will occupy approximately 1U while the platform will occupy approximately 2U. Figure \ref{Figure 3} shows an overview of a satellite and of its payload. Table \ref{Table 1} presents the main characteristics of a satellite.

\begin{table}[h]
\caption{: Main characteristics of each satellite }\label{Table 1}
\begin{center} 
\begin{tabular}{|c|c|}
\hline
    
    Mass & 6-7kg\\
    \hline
    Power consumption & 4W\\
    \hline
    Energy range & 15--200 keV\\
    \hline
    Detectors number for 1 sat. & 7 NaI(Ti) coupled with SiPM\\
    \hline
    Effective geometrical area for 1 sat. & up to 55 cm²\\
    \hline
    Field of view for 1 sat. & 3.46 x $\pi \approx 10.9$ sr total\\
    \hline
    Field of view for 1 sat.(with Earth masking) & 2.73 x $\pi \approx 8.58$ sr total\\
    \hline
    Time resolution & 50 milliseconds\\
    \hline
    Peak count rate per detector & 1500 cts/s/cm²\\
    \hline
    Data rate for 1 satellite & $\leq$ 100 MB/day\\
    \hline

\end{tabular}
\end{center}
\end{table}

The platform provides all services necessary for the mission, mainly the power (solar arrays, batteries and power supply), the telecommunication (S band transmitter and antenna) able to transmit 13 Mbits per second, and the attitude control. The pieces of equipment for the attitude control are not selected yet: it could include a star tracker, a Sun sensor, a magneto-torquer, a GPS receiver, a reaction wheel or a combination of some of them; however there will be no propulsion system. For the demonstration phase, the platform will be designed to allow for a precise attitude and orbit restitution so that we could investigate what will be the main drivers to reach good localising performances.

Given the on-board reduced computing capabilities, we decided not to implement on-board trigger algorithms as for Swift-BAT and SVOM for instance. Rather, the data of each cubesat will be downloaded every 1-3 orbits ($\approx 84$ mn per orbit) to be processed on the ground. The detection and localisation will rely on using the data coming from the whole constellation.  

The science payload is composed of 7 cylindrical NaI(Ti) detectors coupled with SiPM at the bottom, one front-end board for detectors data acquisition, one FPGA (Field Programmable Gate Arrays) board to monitor the payload and one power board to supply power to the payload. The detectors are 80 mm long with a diameter of 12 mm, enveloped in a thin aluminium shell (0.5 mm) for protection against humidity and optical light; they are disposed around a circle of 28.7 mm in diameter, as shown in Figure \ref{Figure 3}. The main shape of the SiPM is a 6 mm side square.
The science payload will record the event counts measured independently by each detector from 15 keV to 200 keV in 5 energy bands every 50 milliseconds. 

\begin{figure}[ht]
\centering
\includegraphics[scale = 0.6]{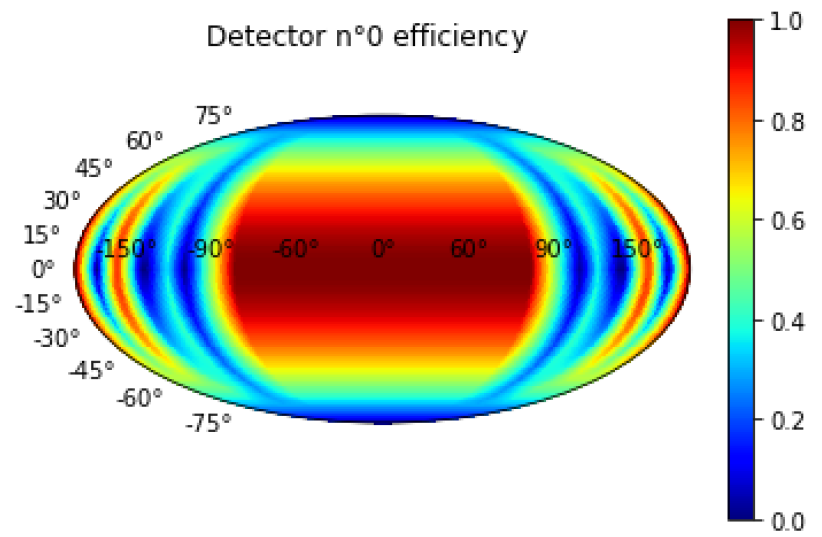}
\caption{: Signal efficiency modulation in azimuth due to the screening of detectors. 
}\label{Figure 4}
\end{figure}

By design, each science payload provides a 1-D localization of detected events. Indeed, the detector cylindrical configuration allows for azimuthal modulation of the total effective area (as shown in Figure \ref{Figure 4} for one detector) due to screening effects between detectors depending on the position of the source with respect to the payload pointing direction. By gathering the 1-D localisation information of every constellation cubesat with different pointing directions, a 2-D sky error box could be reconstructed. 

\section{End-to-end performance simulator}\label{sec3}

In order to optimise the constellation configuration taking into account various constraints (e.g. number of satellites, orbital characteristics, unavailability on the orbit, etc.) to reach our scientific performance requirements, we have built a dynamic simulator. “Dynamic” means that it takes into account the temporal evolution of the constellation configuration. The simulator will also be used along with the data measured with the demonstrator to predict/optimise the performances of the whole constellation. 

In its first version, the simulator only takes into account the geometrical illuminance of the payload detectors by a source located at a certain sky position for a certain configuration of the constellation (i.e. at a certain epoch). In a second version (out of the scope of this paper), we will include a more realistic treatment of the particle--matter interaction using GEANT4 simulation \cite{ref23}.  

We explicit below the inputs and outputs of the simulator as well as the methodology used to build it.

\subsection{Inputs}
\begin{figure}[ht]
\centering
\includegraphics[scale = 0.7]{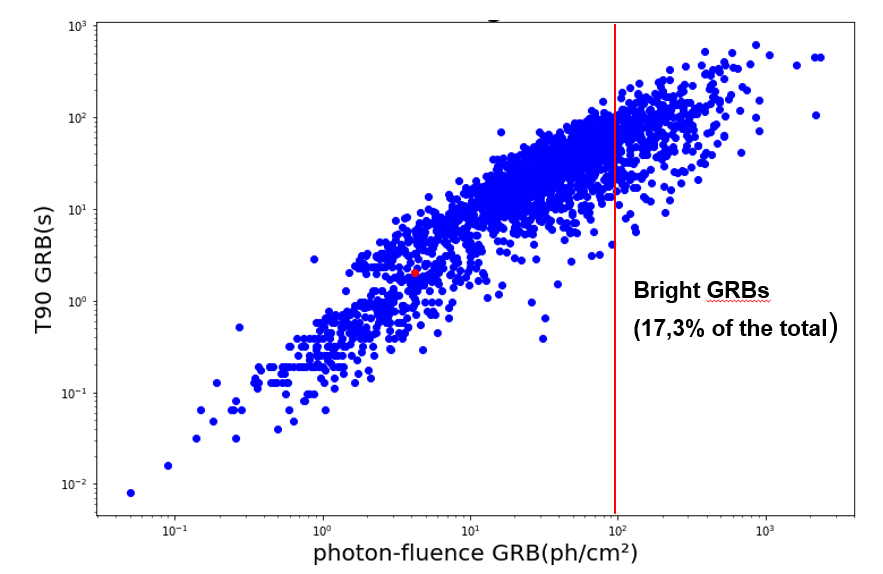}
\caption{: GRB sample from Fermi/GBM catalog used in our simulation in the photon fluence -- $T_{90}$ space. GRB\,170817A is shown as the red point.}\label{Figure 6}
\end{figure}

The two main inputs of the simulator are:\\
1) \underline{The GRB sample}\\
We make use of the Fermi/GBM catalog \cite{ref24}, totalling 2005 GRBs with a large panel of characteristics: the photon fluence in the bandwidth 15--200 keV varies from 0.05 ph/cm$^{2}$ to more than 2000 ph/cm$^{2}$, and $T_{90}$ (time during which 5\% to 95\% of the burst fluence is observed) from 0.1 second to more than 600 seconds. 16.5\% of the GRBs from this sample are short ones. Among the sample, 17\% of the GRBs are "bright GRBs", i.e. GRBs with a photon-fluence larger than 100 ph/cm$^{2}$ in the same bandwidth. 
Figure \ref{Figure 6} displays the GRB sample we used to perform the simulation in a photon fluence vs $T_{90}$ diagram; GRB\,170817A is shown by the red point. In the simulation, we take into account mainly these 2 parameters to estimate the illuminance on the payload detectors. 
The GRB simulated sky positions are given in the J2000 equatorial coordinates system. 

\begin{figure}[ht]
\centering
\includegraphics[scale = 0.8]{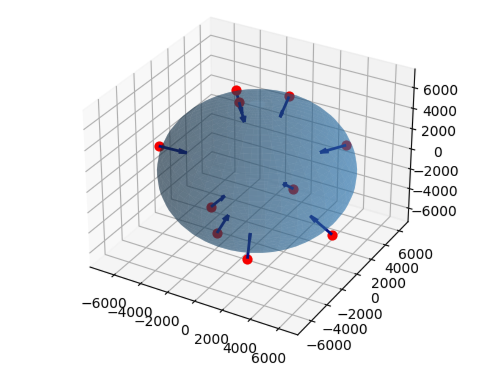}
\caption{: Cartography of 10 satellites (red spots) in the same orbital plane with a geocentric pointing direction (blue arrows). Unity is km.
}\label{Figure 7}
\end{figure}

2)	\underline{The constellation ephemerid}\\
CNES provides ephemerids of a 550 km SSO constellation of 200 satellites distributed into 10 orbital planes with the ascending nodes between 6h and 19h local time (Walker constellation). On each orbital plane, 20 satellites are distributed evenly. The attitude of each satellite is geocentric with yaw-steering orientation (see Section \ref{descrip}).
These ephemerids provide the satellite position, velocity and attitude at 1231 epochs chosen randomly over one year. A subset of the 200 satellites constellation is used for the simulations and, as an example, Figure \ref{Figure 7} shows the position of a 10 cubesat constellation in the same orbital plane around the Earth at a given epoch, with arrows showing the pointing direction of each satellite.

In addition to these two files, we configure the simulation based on a given number of satellites N$_\mathrm{sat}$ (including a satellite failure ratio), a number of GRBs (N) to be considered over N epochs. GRBs can be either chosen randomly from the GRB sample or chosen with specific criteria (for example "only short GRBs" or "a unique GRB"). They are randomly distributed over the whole sky. The epochs could be chosen either randomly over the 1231 values available or fixed to a particular value.

\subsection{GRB detection and localisation}
\label{background}
Once the simulation set-up is done, the simulator then starts the detection likelihood process for each simulated GRB. First, the simulator determines which satellites are located within the SAA or the polar cusps, since in these regions the intense particle background prevents for any detection. This will reduce the observing duty cycle of each satellite and thus of the whole constellation.

\begin{figure}[ht]
\centering
\includegraphics[scale = 0.3]{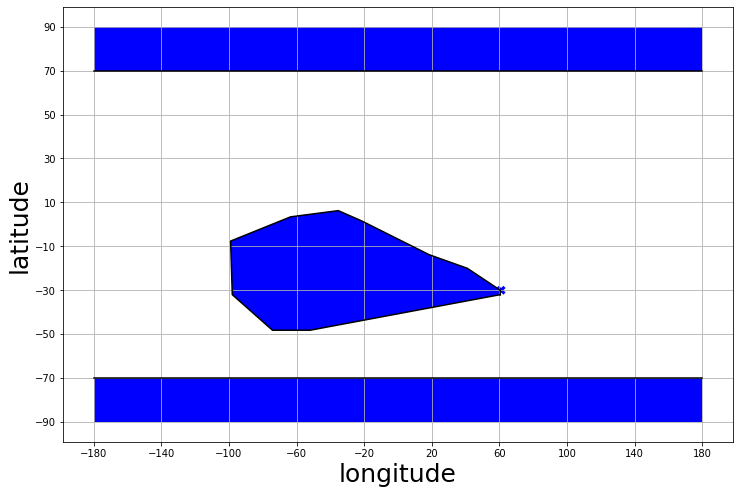}
\caption{ : Regions of satellite unavailability in blue within the SAA and the polar cusps.
}\label{Figure 8}
\end{figure}

The SAA is a region more or less evolving with the Sun activity \cite{ref25}. To get a representative SAA contour for our simulation, we use the mean contour produced by the satellite DMSP (Defense Meteorological Space Program) F18 for the year 2015 \cite{ref25}.
The polar cusps evolve along the day, depending  on the Sun position. However they remain above 70° in absolute latitude value \cite{ref26}. In order to have a conservative assumption, we consider that any satellite being at an absolute latitude value higher than 70° cannot perform any detection.
Figure \ref{Figure 8} shows the SAA and polar cusps taken into account into the simulator.\

Then, the simulator computes the number of counts received from the source and the background noise on each detector of each satellite between 15 keV and 200 keV and over the burst duration $T_{90}$. The background noise consists of 3 main components: the cosmic X-ray background (CXB) \cite{ref28}, the ‘reflection’ component \cite{ref29} produced by the CXB reflection on the Earth atmosphere and the ‘albedo’ component \cite{ref30} produced by the Earth due mainly to the reflection of cosmic rays on the atmosphere. Activation-induced background was not considered at this stage. The main background component is the CXB in the 3U Transat energy bandwidth \cite{ref31}. Thus, as a first approach, we only consider the CXB as the main background source for our simulation.

The simulator computes the signal-to-noise ratio (SNR) between the GRB signal and the noise signal: ratio between the sum of all counts minus the noise received on each detector and the root square of the sum of the noise counts received by each illuminated detector. If the SNR is above the detection threshold fixed arbitrarily at 8, the GRB is considered to be detected.

\begin{figure}[ht]
\centering
\includegraphics[scale = 0.5]{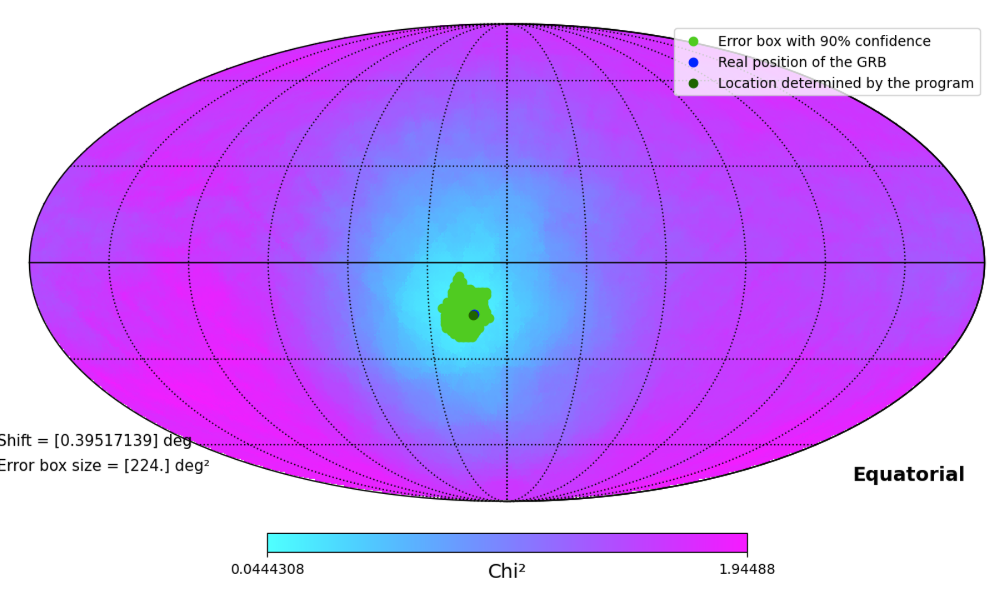}
\caption{: Example of GRB\,170817A detected by a constellation of 200 satellites evenly  distributed over 10 orbital planes. The 90\% c.l. error box is shown in light green (224 deg$^{2}$), the GRB simulated position in blue and the computed one in dark green (with a shift of 0.4 deg).
}\label{Figure 10}
\end{figure}

Then, the simulator performs the localisation of each detected GRB.
To do so, we use an iterative process minimising the $\chi^{2}$ value between the signal produced by any sky position on each detector and the measured counts on the same detectors. The minimum $\chi^{2}$-value provides the computed GRB position. From there, we compute the 90\% confidence level error box (Err90) on this position. Note that the sky meshing used by the simulator is 1 $deg^{2}$ and that we use healpix (Hierarchical Equal Area isoLatitude Pixelization) sky representation \cite{ref32}.\\
Figure \ref{Figure 10} presents a $\chi^{2}$ map showing the simulated position of GRB 170817A (blue dot), the computed one (dark green dot) and its 90\% c.l. error-box (light green area). The localisation is performed with 200 satellites.

\subsection{Outputs}

For each simulated GRB, the simulator provides the source SNR, the number of satellites that could see the burst as well as the total effective area illuminated. In the case of a detection, the simulator also provides the computed position with the corresponding Err90 value. 

\begin{figure}[ht]
\centering
\includegraphics[scale = 0.6]{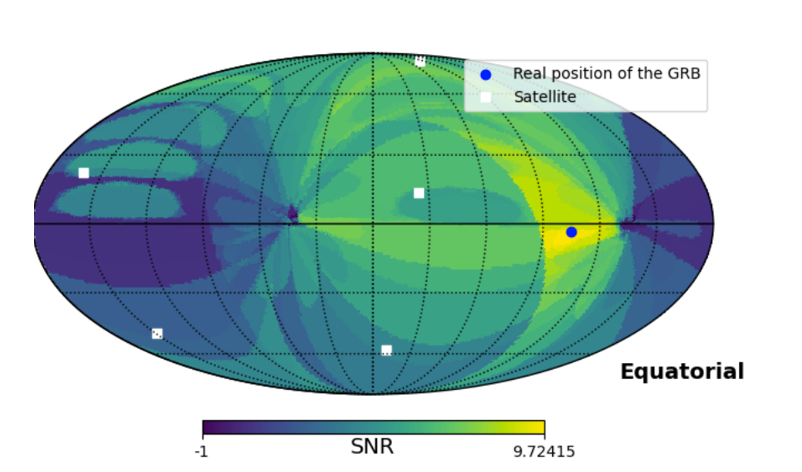}
\caption{ : SNR map for GRB\,090713A and a constellation of 5 satellites evenly distributed on the orbit. The satellites are represented by white squares, while the GRB simulated position is shown as the blue dot. The detection threshold is fixed at SNR $= 8$. 
}\label{Figure 9}
\end{figure}

The simulator also computes for each simulated GRB a SNR or sensitivity map over the whole sky. Figure \ref{Figure 9} shows a sensitivity map for GRB\,090713A and a constellation of five satellites evenly distributed on the orbit. This GRB has a 15 -- 200 keV photon-fluence of 111.4 ph/cm$^2$ and a $T_{90}$-value of 82.8\,s. The satellites are represented by white squares in the figure (four of them were indeed illuminated). The GRB simulated position (the blue dot) is located in a sky region where SNR is above 8. Such a plot allows us to compute the sky fraction over which the burst would have been detected at a given epoch. 
Another use is to reduce the error box size by eliminating the sky regions where the SNR-values are below the detection threshold. This is particularly efficient to reduce the error box sizes of faint GRBs.
For example, regarding the detection of the GRB\,090713A with 5 satellites, the error box size could be reduced by a factor 8.

\section{Performance highlights of 3U Transat}\label{sec4}

\subsection{Evolution of the performances with the number of cubesats}\label{sec4.1}

In order to assess 3U Transat detection and localisation performances, we considered various constellation configurations starting from $N_\mathrm{sat}=3$ (the demonstrator) up to $N_\mathrm{sat}=200$ and a variety of GRBs (short/long, faint/bright bursts) at different epochs.\\ 
Regarding the satellite distribution, a common rule has been applied to ease the comparison between simulations: the satellites are spread on a maximum of 10 orbital planes, with up to 5 satellites evenly distributed per orbital plane for 3, 5, 10, 25 and 50 satellites and with up to 20 satellites per orbit plane for 75, 100, 150 and 200 satellites; it is also considered that there is no satellite failure.\\
This allows to have, as much as possible with the current ephemerids, an even distribution of satellites around the Earth.
The orbits being SSO, the cubesats could go through the polar cusps and the SAA. However, we also studied the case where the polar cusps are not considered in the simulation in order to estimate the impact of these unavailability regions on the performances.

For a given $N_\mathrm{sat}$-value, we chose to consider a sample of $N=300$ GRBs randomly selected within the Fermi/GBM catalogue at 300 different epochs and sky positions. To ease the comparison between simulations at different $N_\mathrm{sat}$-values, we used the same sample of GRBs. 

In order to estimate the dispersion on our results with $N_\mathrm{sat}$, we generated 10 times a new sample of $N=300$ GRBs (with different properties) for $N_\mathrm{sat}=3$ and $N_\mathrm{sat}=100$. 10 trials on the GRB sample appears to be enough to get a reliable estimation of the dispersion of the various computed quantities for a given $N_\mathrm{sat}$-value, while we limited ourselves to the number of $N_\mathrm{sat}$ cases because it would have been time consuming to do it for all cases.\\   

\begin{figure}[ht]
\centering
\includegraphics[scale = 0.3]{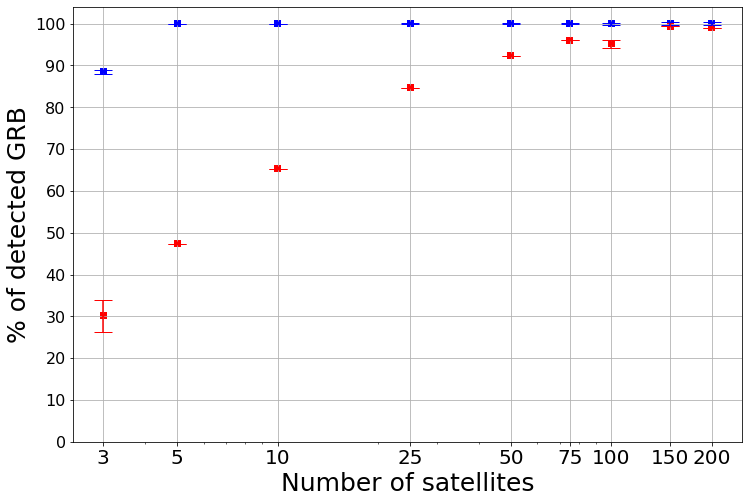}
\caption{: In red, percentage of GRB detection as function of $N_\mathrm{sat}$ for a sample of 300 GRBs. The vertical bars for 3 and 100 cubesats represent the dispersion on the GRB detection percentage computed by generating 10 times the sample of 300 GRBs with different properties. In blue, same information for bright GRBs with 15 -- 200 keV photon fluence $> 100$ ph/cm$^{2}$.   
}\label{Figure 11}
\end{figure}

Regarding the detectability of GRBs (see Figure \ref{Figure 11}), the key performances of 3U Transat are the following: with 100 satellites and more, 3U-Transat is able to detect more than 99\% of a sample of 300 GRBs issued randomly from the Fermi/GBM catalogue. For bright GRBs, the detection performance reaches 100\% for 5 satellites and more.

\begin{figure}[ht]
\centering
\includegraphics[scale=0.245]{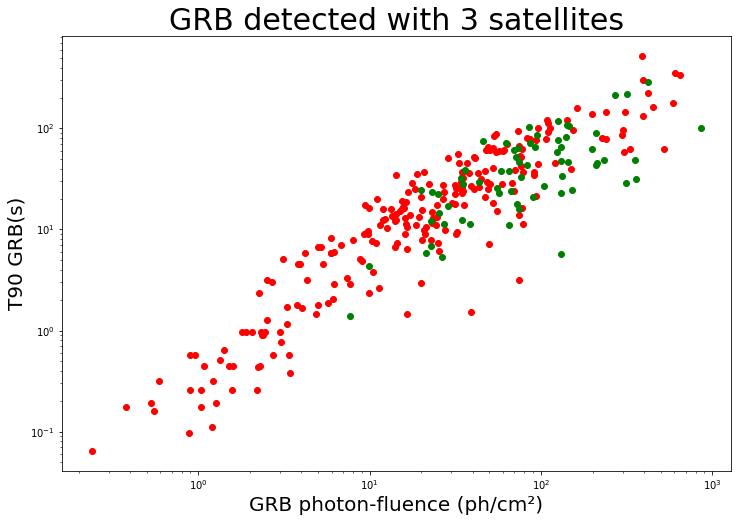}  \includegraphics[scale=0.245]{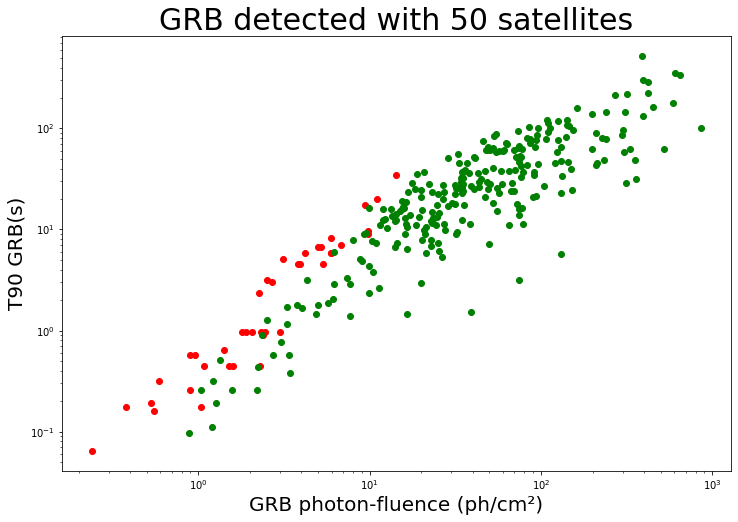} \includegraphics[scale=0.245]{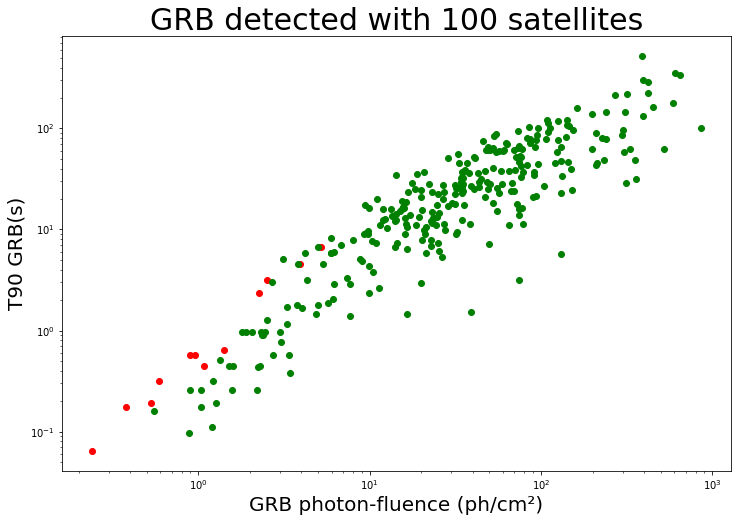} \includegraphics[scale=0.245]{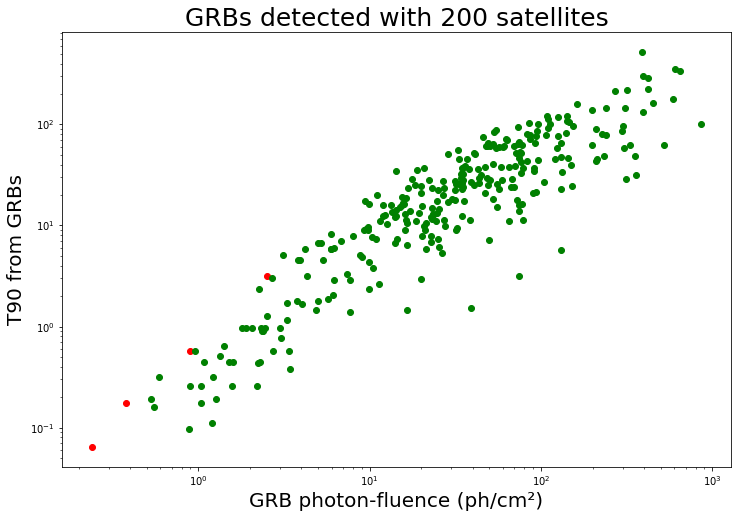} 
\caption{: Evolution of detected (green) and non-detected (red) GRBs in the 15--200 keV photon fluence vs $T_{90}$ space for 3, 50, 100 and 200 satellites. The same sample of 300 GRBs was used in the four plots to ease the comparison of the results. }
\label{Figure 13}
\end{figure}

Figure \ref{Figure 13} shows the detected GRBs in a photon fluence--$T_{90}$ diagram for 3, 50, 100 and 200 satellites. This Figure shows the influence of the 2 main parameters of a GRB on its detectability: as expected, a high fluence improves the detectability. However, a long $T_{90}$ is detrimental to the detectability, since each detector receives in that case more noise photons during the detection.

\begin{figure}[ht]
\centering
\includegraphics[scale = 0.25]{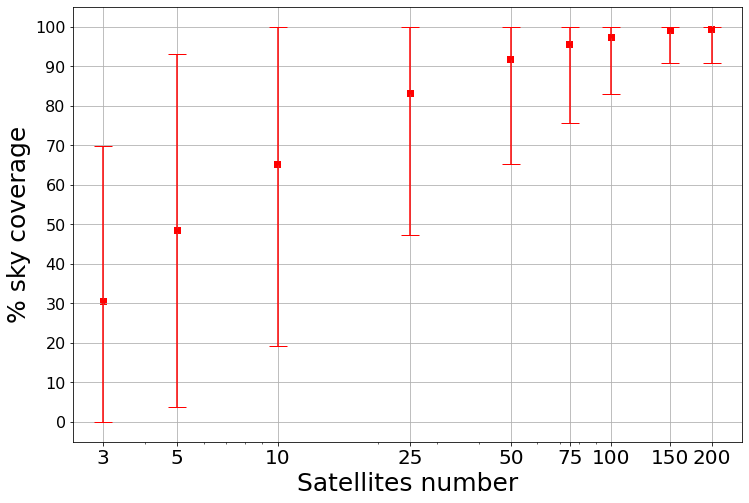}
\includegraphics[scale = 0.25]{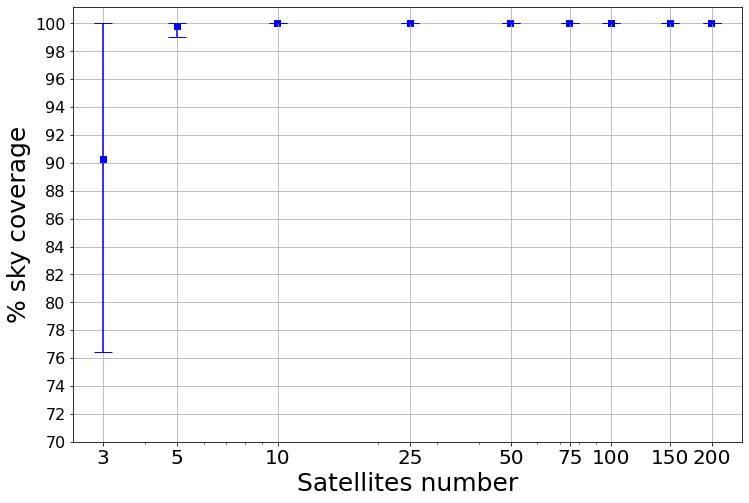}
\caption{: Evolution with N$_\mathrm{sat}$ of the sky fraction over which GRBs were detected with a SNR-value larger than the detection threshold fixed at 8. The squares correspond to the median values of the sky fraction while the vertical bars represent the scatter from the detected GRBs. The figure on the left with red squares is for all GRBs, while on the right it is for bright GRBs only. 
}\label{Figure 14}
\end{figure}

The sky coverage (sky area in which the SNR would be larger than the threshold fixed at 8) is shown in Figure \ref{Figure 14}. It shows that for a sample of 300 GRBs, the sky coverage is larger than 80\% for 100 satellites and more. If we only consider bright GRBs, the sky coverage is close to 100\% for $\geq 5$ satellites.

\begin{figure}[ht]
\centering
\includegraphics[scale = 0.3]{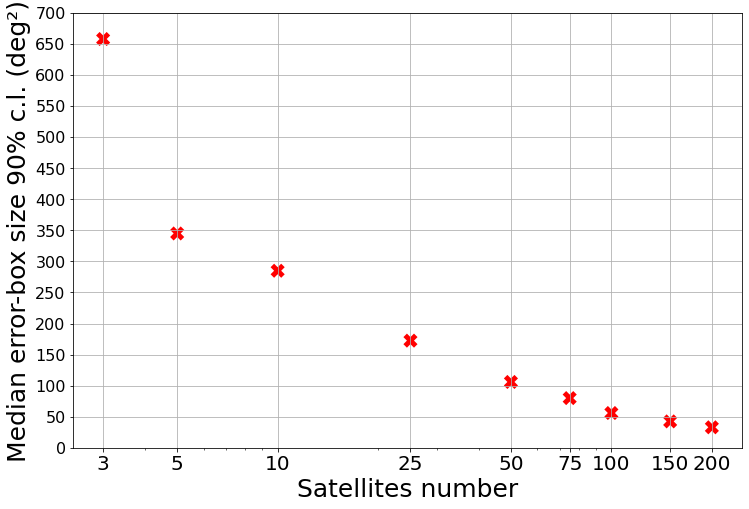}
\caption{: Median Err90 size as function of $N_\mathrm{sat}.$ 
}\label{Figure 16}
\end{figure}

\begin{figure}[ht]
\centering
\includegraphics[scale = 0.25]{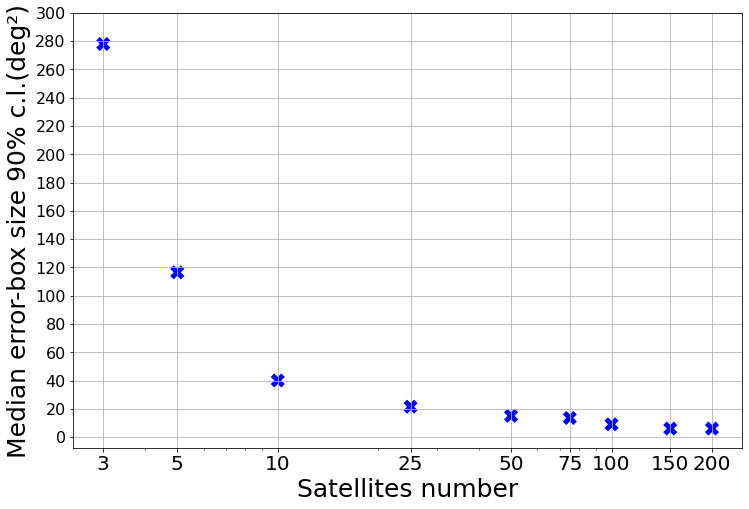} \includegraphics[scale=0.25]{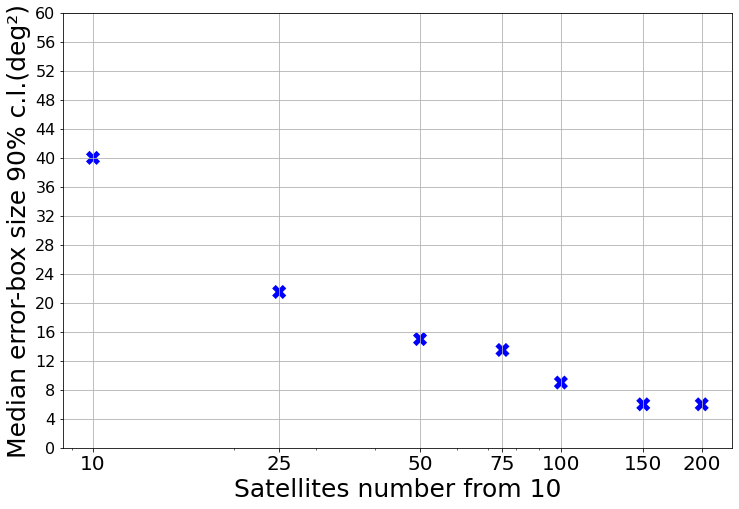}
\caption{: Median Err90 size for bright GRBs. Figure on the left provides the result for 3 to 200 satellites. On the right it is a zoom for 10 to 200 satellites. 
}\label{Figure 17}
\end{figure}

Regarding the localisation, the key performances of 3U Transat are the following (see Figures \ref{Figure 16} and \ref{Figure 17}): with 75 satellites and more, 3U Transat is able to localise a sample of 300 GRBs issued randomly from the Fermi/GBM catalogue with a median 90\% c.l. error box (Err90) size of less than 80 deg$^{2}$ (for 200 satellites, the median Err90 size is equal to 33 deg$^2$). For bright GRBs, the median Err90 size is less than 10 deg$^{2}$ for 100 satellites and more.

\begin{figure}[ht]
\centering
\includegraphics[scale = 0.4]{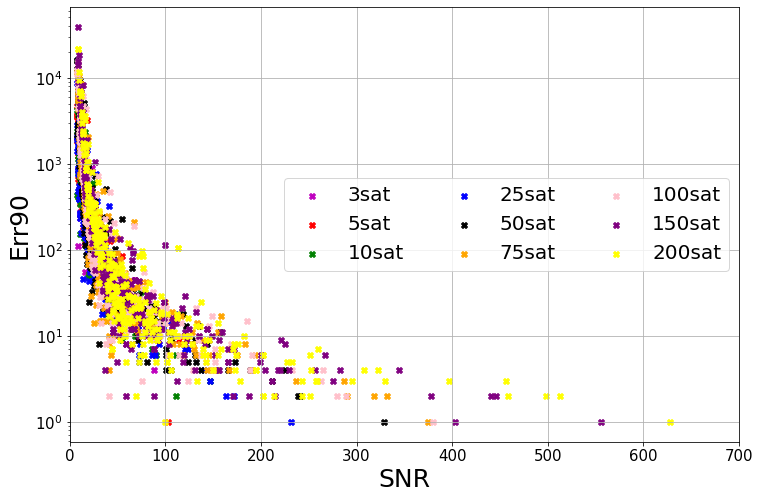}
\caption{: Err90 size as a function of SNR for different constellation sizes (coloured dots).
}

\label{Figure 18}
\end{figure}

Figure \ref{Figure 18} shows the Err90 size evolution with SNR for different constellation sizes. This Figure shows that for the brightest GRBs, the Err90 size could be $\leq 1$ deg$^{2}$. Here, we reach the current simulator sky mesh size of 1 deg$^{2}$. Such GRBs will help us test the ultimate localising performances of 3U Transat.  

\subsection{Focus on GRB\,170817A-like events}\label{sec4.2}

We made a focus on GRBs having similar characteristics (photon fluence and $T_{90}$) than GRB\,170817A with 50, 100 and 200 satellites (Case A). In addition, we considered the case of a GRB with a 15--200 keV photon fluence of 2.1 ph/cm$^{2}$, i.e. half of that of GRB\,170817A (Case B). In that case, we also studied the effect of the satellites passing or not through the polar cusps. In all cases, we simulated 300 different positions at 300 different epochs. \\

\begin{figure}[ht]
\centering
\includegraphics[scale = 0.24]{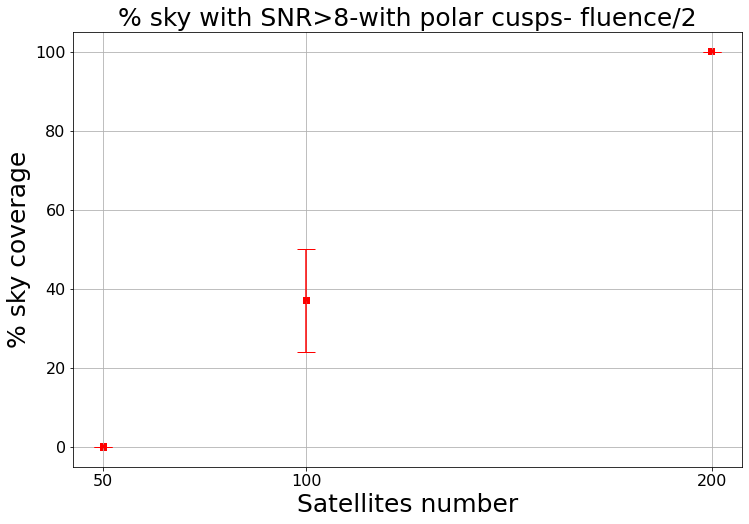}
\includegraphics[scale = 0.24]{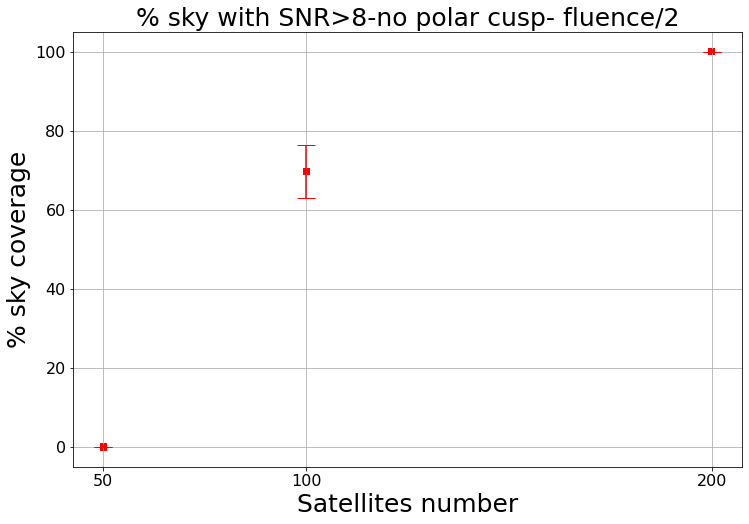}
\caption{: Sky coverage as a function of N$_\mathrm{sat}$ for GRBs with a 15--200 keV photon fluence half of that of GRB\,170817A assuming the satellites pass through the polar cusps (left) or not (right). The vertical bars correspond to the dispersion on the sky coverage from the 300 GRBs simulated at different epochs and sky positions for a given constellation configuration. 
}\label{Figure 19}
\end{figure}

 The results show that in Case A, the sky coverage is 100\% taking into account or not the passing through the polar cusps for $N_\mathrm{sat} \geq 50$ satellites.The median error-box at 90\% c.l. is 570 $deg¨^{2}$ for 100 satellites and 290 $deg^{2}$ for 200 satellites.\\
 In Case B, the results are shown in Figure~\ref{Figure 19} with polar cusps on the left and without polar cusp on the right. It appears that for 50 satellites such GRBs (Case B) could not be detected, while 200 satellites are needed to reach a sky coverage of 100\%. In the latter case, taking into account the polar cusps crossing does not play any significant role, meaning that there are enough satellites available to detect such faint GRBs whatever their sky positions. For smaller N$_\mathrm{sat}$-values, orbits not crossing the polar cusps are more favourable to detect these faint GRBs. Indeed for 100 satellites, the sky coverage is divided by a factor of $\sim 2$ when taking into account the polar cusps crossing.

\subsection{Combination of error-boxes between LVK and 3U Transat}

Our ultimate goal is to combine the Err90-maps from GW detectors (Ligo/Virgo/Kagra) and 3U Transat in order to refine the sky area in which an EM counterpart could be searched for. 
Using our simulator, we investigated such a Err90 sky map combination: to do so, we first selected into the LVK archive an Err90 sky map of a binary NS (BNS) event; then within this error box, we simulated numerous short GRBs (typically 300) from our GRB sample; for the detected events, we computed their Err90 sky map; finally, we computed the common sky area between the LVK \& 3U Transat Err90 sky maps for each detected GRB.

\begin{figure}[ht]
\centering
\includegraphics[scale = 0.6]{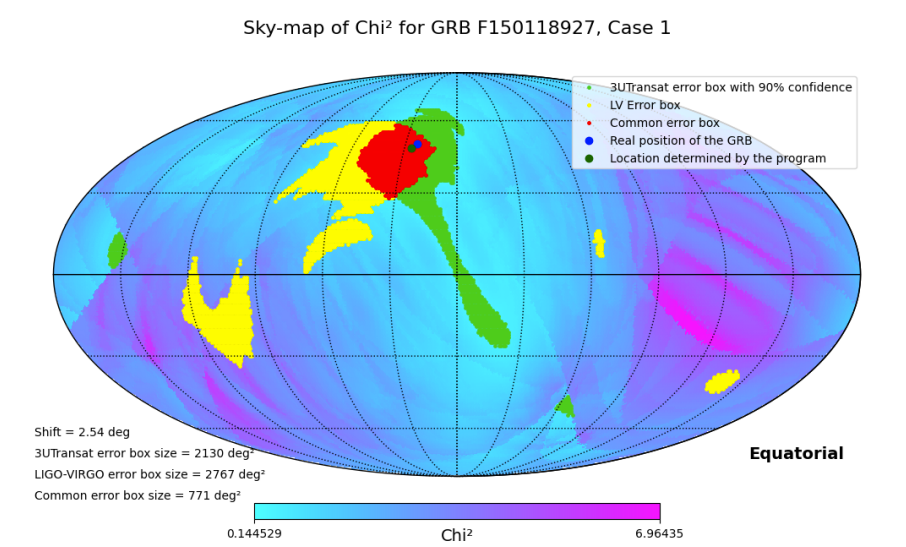}
\caption{: Combination of LVK and 3U Transat Err90 sky maps.}
\label{Figure 24}
\end{figure}

The chosen event from LVK is the BNS event ID S190822c\cite{ref33} with an Err90 size of 2767 deg$^{2}$.
Figure \ref{Figure 24} shows an example of common sky area (in red) between LVK and 3U-Transat: the LVK Err90 is shown in yellow, while the 3U Transat one in green. For this simulation the 3U Transat constellation was made of 10 satellites evenly distributed over one orbit. On the 300 short GRBs simulated within the LVK error box, 36 were indeed detected.
The median common area between LVK and 3U Transat is then equal to 441.6 deg$^{2}$, which represents a reduction by a factor $\approx 6$ of the LVK initial error box. 

\section{Discussion and conclusion}\label{sec5}

3U Transat is a French cubesat constellation project with the aim to survey all the high energy sky in the 15-200 keV band to detect GRBs of all types (including local ($<$ 60 Mpc) underluminous GRB\,170817A-like events) and other high energy transients in synergy with MM facilities, starting from mid-2027 with a 3 yr demonstrator (3 cubesats). Every cubesat is identical embarking on a 3U format a 1U science payload providing 1-D localisation performances. Gathering localisation information from the constellation cubesats with varied attitude allows to reconstruct a 2-D sky error box for the detected events. We recall that no on-board trigger software will be embarked on the cubesats. Instead, all the data from the constellation will be regularly  dumped to the ground for detection and localisation purposes, enabling to broadcast alerts to the community in less than 4--5\,h after data collection on-board. This delay shall be enough to search for an off-axis GRB afterglow and/or a kilonova associated to a BNS merger.  

In this paper, we presented our geometrical and dynamic simulator that we used to investigate the constellation configurations (e.g. number of cubesats, orbits and attitude, etc.) that could fulfil our science requirements.\\

Thanks to the results presented in Sections \ref{sec4.1} and \ref{sec4.2}, we conclude that 3U Transat will be able to reach most of our science requirements if the number of satellites is equal or greater than 100. Bright GRBs with photon fluence $F > 10^2$ ph/cm$^2$ in the 15--200 keV band could be used to measure the ultimate performances of our localisation principle. Indeed, simulation showed that Err90 sizes could be less than 1 deg$^2$. We also showed that to have an all-sky view of GRB\,17087A-like or twice fainter GRBs it is necessary to deploy either 200 cubesats spread over different orbital SSO-like planes or to choose for less satellites more adequate orbits avoiding the polar cusps and SAA crossing. Note that the computation of the GRB detection rate as a function of N$_\mathrm{sat}$ is out of the scope of the present paper.\\

Given our assumptions to build the geometrical and dynamic simulator, it presents some limitations to fully estimate 3U Transat performances for the whole constellation. For instance, the detection process is not done the way we expect to do it i.e. making use of all the lightcurves measured on-board to search for coincident temporal excesses in different energy bands and timescales. Generating more realistic output data from each science payload is mandatory to develop our data reduction and analysis tools as well as to investigate the best way to combine them with data coming from other missions. 

To do so, we are currently developing a physically-based dynamic mission simulator with a refined treatment of both the background noise and the GRB properties. This new simulator takes the photon--matter interaction as well as material transparency with energy into account. To compute the three main components of the background noise (CXB, albedo and reflection) in the 15--200 keV energy band (see Section~\ref{background}), we use the Particle Interaction Recycling Approach (PIRA -- \cite{ref34}), a method first introduced to compute the SVOM/ECLAIRs dynamic background. This method relies on performing Monte-Carlo GEANT4-based simulation of the various background components to build a large database of events that could be used to reconstruct the background rate at any location on the orbit.
 Regarding GRB simulation, we will make use of GRB lightcurves and spectra from Swift and Fermi/GBM GRB catalogues. We plan to use a ray-tracing method to do so, as done for the SVOM mission \cite{ref34}.           

Once ready, this physical simulator will allow us to investigate how to set some parameters (e.g. the SNR threshold, the choice of the energy bandwidths to record the counting) to optimise the constellation performances. We will also estimate the expected GRB detection rate as function of $N_\mathrm{sat}$.

Regarding the localisation, we also plan to investigate other ways to do it. We may be able to use the triangulation method based on the measure of time delays between different satellites \cite{ref35}. However, this technique may be usable only for bright GRBs given that the satellites will be located in LEO (implying small time delays over which to look for count excesses above the background). 

Finally the demonstration phase is key to validate the 3U Transat measuring principle, to measure the background level and evolution along the orbit as well as to investigate the impacts of the ageing of SiPM on the constellation performances. Moreover, the demonstrator data along with our simulator will be helpful to predict the performances of a more ambitious constellation.

\acknowledgments 
 
We are grateful to the CNES French Space Agency for providing us with the ephemerids used to perform the simulation and for their support on the 3U Transat project.

\clearpage 



\end{document}